\documentclass[]{article}

\usepackage{graphics}
\usepackage{graphicx}
\usepackage[linesnumbered,ruled,vlined]{algorithm2e}
\usepackage{algorithmicx}
\usepackage{algpseudocode}
\usepackage[]{algorithm2e}
\usepackage{amsfonts}
\usepackage{amsmath}
\usepackage{bm}
\usepackage{bbm}
\usepackage{array}
\usepackage{breqn}
\usepackage{epstopdf}
\usepackage{booktabs}
\usepackage{natbib}
\usepackage{siunitx}

\usepackage{adjustbox}
\usepackage{multirow}
\usepackage{color}

\graphicspath{{./Figures/}}

\newcolumntype{d}{>{\displaystyle}c}
\newcommand{\ts}{\textsuperscript}

\definecolor{red}{rgb}{0, 0, 0}
\usepackage{subfigure}

\usepackage{authblk}
\title{Improvement of PolSAR Decomposition Scattering Powers Using a Relative Decorrelation Measure}
\author[1]{D.Ratha}
\author[1]{M.Surender}
\author[1]{A.Bhattacharya}
\affil[1]{Centre of Studies in Resources Engineering, Indian Institute of Technology Bombay, Powai, Mumbai-400076, India}
\date{}                     
\begin{document}
%

\maketitle

\begin{abstract}
In this letter, a methodology is proposed to improve the scattering powers obtained from model-based decomposition using Polarimetric Synthetic Aperture Radar (PolSAR) data. The novelty of this approach lies in utilizing the intrinsic information in the off-diagonal elements of the 3$\times$3 coherency matrix $\mathbf{T}$ represented in the form of complex correlation coefficients. Two complex correlation coefficients are computed between co-polarization and cross-polarization components of the Pauli scattering vector. The difference between modulus of complex correlation coefficients corresponding to $\mathbf{T}^{\mathrm{opt}}$ (i.e. the degree of polarization (DOP) optimized coherency matrix), and $\mathbf{T}$ (original) matrices is obtained. Then a suitable scaling is performed using fractions \emph{i.e.,} $(T_{ii}^{\mathrm{opt}}/\sum\limits_{i=1}^{3}T_{ii}^{\mathrm{opt}})$ obtained from the diagonal elements of the $\mathbf{T}^{\mathrm{opt}}$ matrix. Thereafter, these new quantities are used in modifying the Yamaguchi 4-component scattering powers obtained from $\mathbf{T}^{\mathrm{opt}}$. To corroborate the fact that these quantities have physical relevance, a quantitative analysis of these for the L-band AIRSAR San Francisco and the L-band Kyoto images is illustrated. Finally, the scattering powers obtained from the proposed methodology are compared with the corresponding powers obtained from the \textcolor{red}{Yamaguchi \emph{et. al.,} 4-component} (Y4O) decomposition and the \textcolor{red}{Yamaguchi \emph{et. al.,} 4-component Rotated} (Y4R) decomposition for the same data sets. The proportion of negative power pixels is also computed. The results show an improvement on all these attributes by using the proposed methodology.
\end{abstract}

\section{Introduction}
There are a number of \textcolor{red}{Polarimetric Synthetic Aperture Radar} (PolSAR) scattering power decompositions based on physical scattering models which have been utilized for target classification, detection and geophysical parameter extraction. The Freeman-Durden 3-component decomposition (FDD)~\citep{freeman98} and the Yamaguchi \emph{et. al.,} 4-component (Y4O) decomposition~\citep{Yamaguchi2005} have been extensively used in applications since its inception.  

The target coherency matrix, $\mathbf{T}$ which encompasses second-order statistical information of incoherent scattering phenomenon is used in PolSAR data analysis. The off-diagonal terms, especially the correlations between co-polarization and cross-polarization terms provide direct information about polarization orientation angle induced by terrain azimuth slopes~\citep{lee2002estimation}. Moreover, from statistical perspective, they are also correlation quantities between various kind of elementary scattering mechanisms. Hence, it becomes imperative to use these quantities to quantify various PolSAR target scattering characteristics~\citep{ainsworth2008polarimetric,chen2015crop,frery2014contrast}. 

In the context of advances in model-based decomposition studies, the original $\mathbf{T}$ matrix is transformed by real and complex rotation before applying the \textcolor{red}{Yamaguchi \emph{et. al.,} 4-component Rotated} (Y4R) decomposition~\citep{YAMAGUCHI2011} and the \textcolor{red}{Generalized 4-component Unitary} (G4U) decomposition~\citep{singh13} respectively. Furthermore, the \textcolor{red}{Adaptive Generalized 4-component Unitary} (AG4U) decomposition~\citep{Bhattacharya15_AG4U} decomposition introduced certain complex rotation matrices to transform the original coherency matrix $\mathbf{T}$, whereas the \textcolor{red}{Stochastic Distance based Y4O} (SD--Y4O)~\citep{Bhattacharya15_SD_Y4O} decomposition is based on statistical information theory. Recently, a methodology $(\mathrm{AGU-DOP})$ has been proposed to enhance target information in PolSAR data using a criteria on the Degree of Polarization ($\mathrm{DOP}$)~\citep{Bhattacharya15_AG4U}. 

\textcolor{red}{Few existing scattering power decomposition methods utilize some of the off-diagonal elements of the coherency (or covariance) matrix. These off-diagonal elements provide the correlation between the components of the Pauli scattering vector. The novelty of this technique is realized by utilizing the change in the off-diagonal elements (or correlation quantities) under the AGU--DOP transformation with respect to the original coherency matrix for the modification of the scattering powers.}

\textcolor{red}{The results show that the overall even-bounce powers over rotated urban areas have significantly improved with the reduction of volume powers. The proportion of pixels with negative powers have also decreased from the Y4O decomposition. The proposed method is both qualitatively and quantitatively compared with the results obtained from the Y4O and the Y4R decompositions on the original $\mathbf{T}$.}

\section{Methodology}
\subsection{Transformation of Coherency Matrix}
\textcolor{red}{In PolSAR, the scattering matrix $\mathbf{[S]}$ which is expressed in the backscatter alignment (BSA) convention is given in the linear horizontal (H) and linear vertical (V) polarization basis as,
	\begin{equation}
	\mathbf{[S]}=
	\left[\begin{array}{cc}
	S_{\mathrm{HH}} & S_{\mathrm{HV}}\\
	S_{\mathrm{VH}} & S_{\mathrm{VV}}
	\end{array}\right] 
	\label{Eq:scattering_matrix}
	\end{equation}
	Assuming reciprocity ($S_{\mathrm{HV}} = S_{\mathrm{VH}}$), the $3\times3$ coherency matrix $\mathbf{T}$ is obtained from the Pauli scattering vector $\textit{\textbf{k}}_{\mathrm{p}} = \frac{1}{\sqrt{2}}[S_{\mathrm{HH}}+S_{\mathrm{VV}}, S_{\mathrm{HH}} - S_{\mathrm{VV}},2S_{\mathrm{HV}}]^{\mathrm{T}}$ (where the subscript $\mathrm{H}$ and $\mathrm{V}$ denotes horizontal and vertical polarized waves and the superscript $\mathrm{T}$ denotes transpose) as,}
\begin{equation}
	\mathbf{T}  = \langle \textit{\textbf{k}}_{\mathrm{p}} \textit{\textbf{k}}_{\mathrm{p}}^* \rangle = \left[\begin{array}{lcl}
		T_{11} & T_{12} & T_{13} \\
		T_{21} & T_{22} & T_{23} \\
		T_{31} & T_{32} & T_{33} 
	\end{array}\right] \label{coh}
\end{equation} 
\textcolor{red}{where $\langle \: \rangle$ denotes the ensemble average, the superscript $*$ denotes the complex conjugate transpose and the subscript $\mathrm{p}$ denotes the Pauli basis. A $3\times3$ Hermitian matrix $\mathbf{U}$ (i.e $\mathbf{U}^*\mathbf{U} = \mathbf{U}\mathbf{U}^* = \mathbf{I}$) can be used to modify the original coherency matrix into,}
\begin{equation}
\mathbf{T}(\mathbf{U}) = \mathbf{U}\mathbf{T}\mathbf{U}^*= \langle (\mathbf{U}\;\textit{\textbf{k}}_{\mathrm{p}}) (\mathbf{U}\;\textit{\textbf{k}}_{\mathrm{p}})^* \rangle
\label{eqn:inprt}
\end{equation}
as $\langle \: \rangle$ preserves linear operations. Hence, multiplication by a Hermitian matrix can be interpreted as the modification of the $\textit{\textbf{k}}_{\mathrm{p}}$ to obtain a modified coherency matrix. Similar transformations of the coherency matrix, where \textcolor{red}{$\mathbf{U} \in \mathrm{SU(3)}$ (group of $3\times3$ special unitary matrices)} have been performed in PolSAR scattering power decomposition study~\citep{YAMAGUCHI2011,singh13, Bhattacharya15_AG4U}. In radar polarimetry, the $\mathrm{SU(3)}$ group is related to the polarization basis transformation for the coherency matrix. Due to the orientation of urban structures about the radar line of sight and non-zero azimuthal slopes~\citep{schuler1999compensation,lee2002estimation}, the observed $\textit{\textbf{k}}_{\mathrm{p}}$ needs to be modified by suitable transformations to correct these factors that inherently alter the perceived scattering characteristics of the targets~\citep{Lee2011}. 

Under the polarization basis transformation, the Degree of Polarization ($\mathrm{DOP}$) parameter is an important criterion in estimating $\mathbf{T}^{\mathrm{opt}}$~\citep{Bhattacharya_2016RSL}. \textcolor{red}{The coherency matrix $\mathbf{T}^{\mathrm{opt}}$ is obtained after the AGU-DOP transformation on $\mathbf{T}$.} The $\mathbf{T}^{\mathrm{opt}}$ is then used for target characterization~\citep{Bhattacharya_2016RSL}, enhanced target decomposition scattering powers~\citep{Bhattacharya15_AG4U,Bhattacharya_2016RSL} and snow surface dielectric constant estimation~\citep{Manickam_2106JSTARS}. In this study, we utilize the information content in the off-diagonal elements of the $\mathbf{T}^{\mathrm{opt}}$ to improve the scattering powers of an existing decomposition.

\subsection{Relative decorrelation measure from co- and cross-polarization channels}
The coherency matrix $\mathbf{T}$ contains the second-order statistics of the backscattering phenomenon. It is commonly assumed that the components of $\textit{\textbf{k}}_{\mathrm{p}}$ follow a central Gaussian distribution with expectation to be zero. Thus, the off-diagonal elements of $\mathbf{T}$ correspond to the covariance and the diagonal elements to the variance of these components. \textcolor{red}{The expression for the polarization orientation angle (POA) $\theta$ induced by azimuth terrain slopes contains the real part ($\mathrm{Re}(\cdot)$) of the $T_{23} = \langle (S_{\mathrm{HH}} - S_{\mathrm{VV}}) S_{\mathrm{HV}}^* \rangle$ term, whereas the corresponding imaginary part contains information on returns from helical targets as given in~\citep{krogager95z}. The orientation compensation affects more the $\mathrm{Re}(\langle (S_{\mathrm{HH}} - S_{\mathrm{VV}}) S_{\mathrm{HV}}^* \rangle)$ than the $T_{13} = \langle (S_{\mathrm{HH}} + S_{\mathrm{VV}}) S_{\mathrm{HV}}^* \rangle$ term~\citep{lee2002estimation}. The phase of $\langle (S_{\mathrm{HH}} + S_{\mathrm{VV}}) S_{\mathrm{HV}}^* \rangle$ depends weakly on the orientation angle. The rotated $T_{13}$ values are $T_{12}\sin(2\theta)+T_{13}\cos(2\theta)$.}

In a reflection symmetric medium, the co- and the cross-polarized coupling terms are absent \emph{i.e.,} $T_{13} = T_{23} = 0$. But for a natural target both are never zero simultaneously for any value of $\theta$. The compensation of $\mathbf{T}$ by $\theta$ makes the $\mathbf{Re}(T_{23}) = 0$ while the $T_{13}$ element is slightly reduced. However, a multi-looking process tends to reduce the magnitude of the orientation compensated $T_{13}$ element~\citep{lee2002estimation}. In~\cite{YAMAGUCHI2011} decomposition, the POA compensation is applied prior to the decomposition to make $\mathbf{Re}(T_{23}) = 0$ reducing the number of polarimetric parameters from 9 to 8. Moreover, like FDD~\citep{freeman98}, the $T_{13}$ element is unused in the Y4O~\citep{Yamaguchi2005,YAMAGUCHI2011}. However, a recent model-based scattering power decomposition~\citep{Arii11} considers the $T_{13}$ element which is theoretically made zero by a proper POA compensation.      

Therefore, the off-diagonal elements $(\emph{i.e.,} \;T_{13} \;\mathrm{and}\; T_{23})$ which are representative of the correlations between the co- and the cross-polarized return measurements should be investigated and suitably taken into account while assessing the scattering information from PolSAR data. In this study we have analyzed the complex correlation coefficients $\rho_{13}$ and $\rho_{23}$ defined as,
\begin{subequations}
	\begin{align}
	\rho_{13} &= \frac{T_{13}}{\sqrt{T_{11}T_{33}}}\label{rho13}\\
	\rho_{23} &= \frac{T_{23}}{\sqrt{T_{22}T_{33}}}.\label{rho23}
	\end{align}
\end{subequations} 
Here $\rho_{13}$ corresponds to the coupling between the $1\ts{st}$ (surface) and the $3\ts{rd}$ (volume) component of the $\textit{\textbf{k}}_{\mathrm{p}}$ whereas $\rho_{23}$ corresponds to the coupling between the $2\ts{nd}$ (double-bounce) and the $3\ts{rd}$ (volume) component of the $\textit{\textbf{k}}_{\mathrm{p}}$. The magnitude of the correlation $(\vert\rho_{ij} \vert)$ is in $[0,1]$. 

In the context of this study, the magnitude of the correlation coefficients $\rho_{13}$ and $\rho_{23}$ obtained from $\mathbf{T}$ is compared with that of $\rho_{13}^{\mathrm{opt}}$ and $\rho_{23}^{\mathrm{opt}}$ obtained from $\mathbf{T}^{\mathrm{opt}}$~\citep{Bhattacharya_2016RSL} as,
\begin{subequations}
	\begin{align}
	\delta_{13} &= (\vert \rho_{13} \vert - \vert \rho_{13}^{\mathrm{opt}} \vert)\mathbbm{1}_{[0,1]} \\
	\delta_{23} &= (\vert \rho_{23}\vert  - \vert \rho_{23}^{\mathrm{opt}}\vert)\mathbbm{1}_{[0,1]} 
	\end{align}
\end{subequations}
where $\mathbbm{1}_{[0,1]}$ is the characteristic function of $[0,1]$ interval. Thus, larger the value of $\delta_{ij}$, smaller is the magnitude of the co- and the cross-polarized correlation $(\emph{i.e.,} \;T_{13}^{\mathrm{opt}} \;\mathrm{and}\; T_{23}^{\mathrm{opt}})$ in $\mathbf{T}^{\mathrm{opt}}$ compared to the original $\mathbf{T}$. This attribute is clearly shown in~\citep{Bhattacharya_2016RSL} using a polarimetric similarity parameter. The optimized Mueller matrix $\mathbf{M}^{\mathrm{opt}}$ (obtained as a one-to-one mapping from the $\mathbf{T}^{\mathrm{opt}}$ matrix) has been shown to be closer to an elementary target than the original Mueller matrix $\mathbf{M}$ obtained similarly. As the $\mathbf{T}^{\mathrm{opt}}$ is obtained by optimizing the effective $\mathrm{DOP}$, it is expected that $\mathbf{T}^{\mathrm{opt}}$ lowers the correlation between the co- and the cross-polarized returns. More than $90$\% of the total number of pixels in the images analyzed in this study show this trend. But, an oscillation of the $\delta_{ij}$ parameter about zero may be present as the $\mathrm{AGU-DOP}$ methodology does not explicitly use the off-diagonal elements. Hence, only the pixels with non-negative correlation differences (i.e. $\delta_{ij} \ge 0$) have been considered for modifying the scattering powers.

\subsection{Improvement of the Yamaguchi et. al., 4-component scattering powers}
\begin{figure}[!hbt]
	\centering
	\includegraphics[width=0.5\linewidth,keepaspectratio=true]{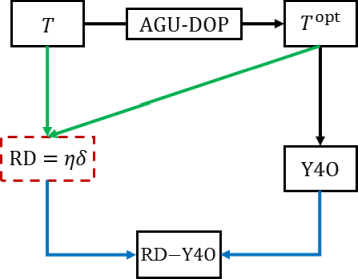}
	\caption{Flowchart of the proposed methodology}
	\label{fig:flowchart}
\end{figure}

In this section, a methodology is proposed to modify the Yamaguchi \emph{et. al.} 4-component scattering powers decomposition (Y4O) derived from the $\mathbf{T}^{\mathrm{opt}}$. The analysis of the off-diagonal elements (the correlations between the co- and the cross-polarized components) of the coherency matrix in terms of the $\delta_{ij}$s are used to compensate the scattering powers. \textcolor{red}{The $\delta_{ij}$s do not give information about the dominant component.} Hence, it is appropriate to use the fractions $\eta_{1} = T_{11}^{\mathrm{opt}}/(\mathrm{span})$, and $\eta_{2} = T_{22}^{\mathrm{opt}}/(\mathrm{span})$, where $\mathrm{span}=T_{11}^{\mathrm{opt}}+T_{22}^{\mathrm{opt}}+T_{33}^{\mathrm{opt}}$. A relative decorrelation (RD) measure defined as $\zeta_{i}=\eta_{i}\delta_{ij}$ is suitably used to compensate the Y4O scattering powers. \textcolor{red}{The flowchart of the proposed methodology is shown in Figure~\ref{fig:flowchart}. The original $\mathbf{T}$ is first transformed to $\mathbf{T}^{\mathrm{opt}}$ under the $\mathrm{AGU-DOP}$. The $\eta_{i}$s are computed from $\mathbf{T}^{\mathrm{opt}}$ along with the Y4O scattering powers. Then the $\delta_{ij}$s are computed using both $\mathbf{T}$ and $\mathbf{T}^{\mathrm{opt}}$. After that the $\zeta_{i}$s are formed from the product of $\delta_{ij}$ and $\eta_{i}$ which is termed as the relative decorrelation (RD). Finally the Y4O powers obtained from $\mathbf{T}^{\mathrm{opt}}$ are modified to RD--Y4O powers as given in Equation~\eqref{eqn:RD-Y4O}(a)--(c).}
\begin{subequations}
	\begin{align}
	\overset{\sim}{{P}}_{\mathrm{s}} &=  P_{\mathrm{s}}^{\mathrm{opt}} + \zeta_{1}P_{\mathrm{v}}^{\mathrm{opt}}, \\
	\overset{\sim}{{P}}_{\mathrm{d}} &=  P_{\mathrm{d}}^{\mathrm{opt}} + \zeta_{2}P_{\mathrm{v}}^{\mathrm{opt}},\\ 
	\overset{\sim}{{P}}_{\mathrm{v}} &=  [1 - (\zeta_{1} + \zeta_{2})]P_{\mathrm{v}}^{\mathrm{opt}}
	\end{align}
	\label{eqn:RD-Y4O}
\end{subequations}
where $\overset{\sim}{{P}}_{\mathrm{s}}$, $\overset{\sim}{{P}}_{\mathrm{d}}$ and $\overset{\sim}{{P}}_{\mathrm{v}}$ are the improved (RD--Y4O) surface, double bounce, and volume scattering powers respectively. The $P_{\mathrm{s}}^{\mathrm{opt}}$, $P_{\mathrm{d}}^{\mathrm{opt}}$, and $P_{\mathrm{v}}^{\mathrm{opt}}$ are the corresponding Y4O scattering powers obtained from $\mathbf{T}^{\mathrm{opt}}$. 

\section{Data sets and study area}
In this study, we have used a 4-look AIRSAR L-band polarimetric SAR data over San Francisco, USA with a ground resolution of 12~m and an ALOS-2 L-band polarimetric SAR data over Kyoto, Japan with a ground resolution of 15~m. The ALOS-2 image is multi-looked with a factor of 3 in range and 5 in azimuth. Buildings oriented about radar LOS with azimuth and range slopes is a common feature in both these areas. A window size of 3$\times$3 is used to obtain the decomposition powers.
 
\section{Results and discussion}
\begin{figure}[!hbt]
	\centering
		\subfigure[]{\adjincludegraphics[trim={{0.03\width} 0 {0.13\width} 0},clip, height=0.31\textwidth]{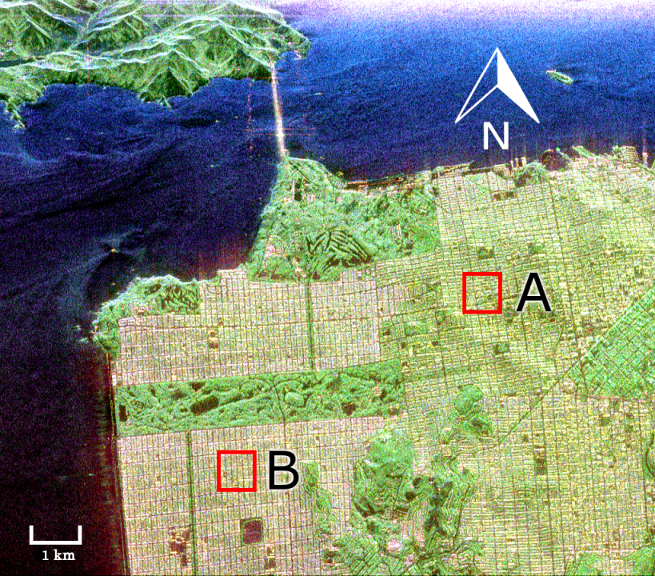}} 
		\subfigure[]{\includegraphics[height=0.31\textwidth]{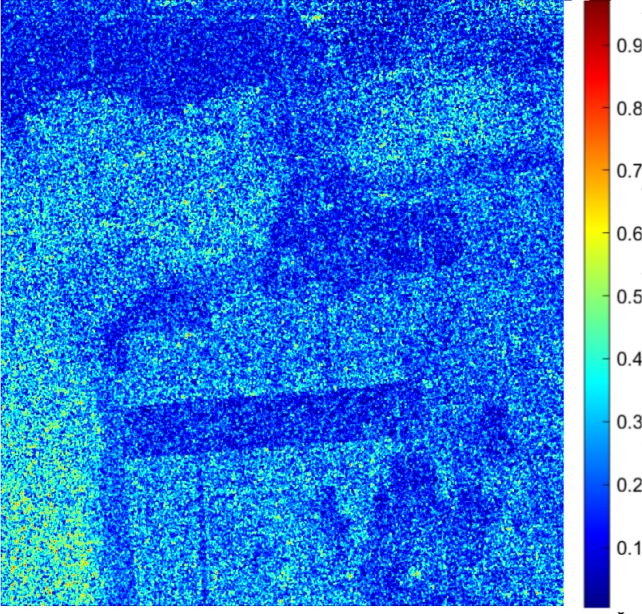}} 
		\subfigure[]{\includegraphics[height=0.31\textwidth]{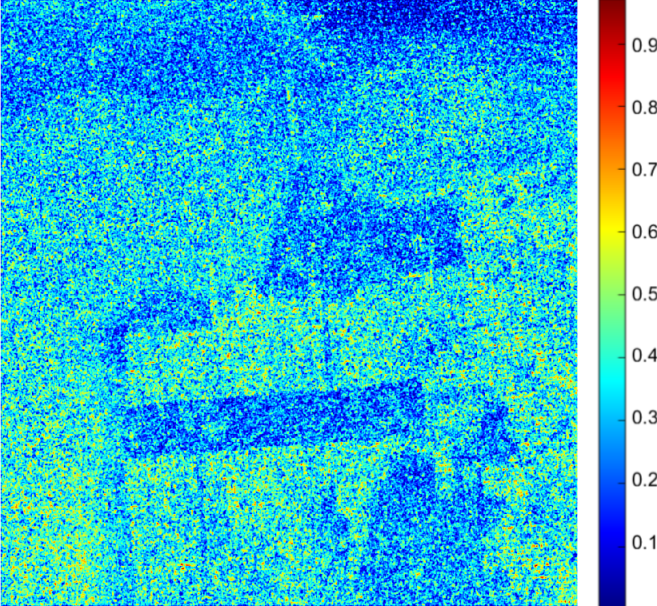}} \\
		\subfigure[]{\adjincludegraphics[trim={0 {0.6\width} 0 0},clip, height=0.30\textwidth]{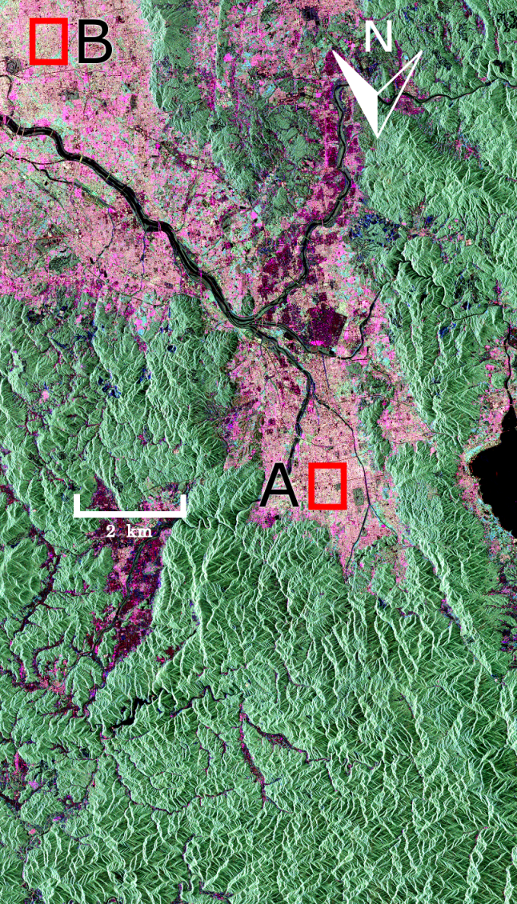}} 
		\subfigure[]{\adjincludegraphics[trim={0 {0.33\width} 0 0},clip, height=0.30\textwidth]{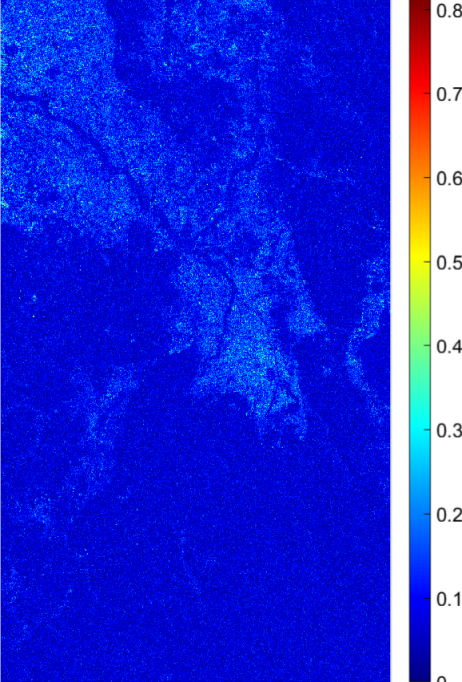}}  
		\subfigure[]{\adjincludegraphics[trim={0 {0.344\width} 0 0},clip, height=0.30\textwidth]{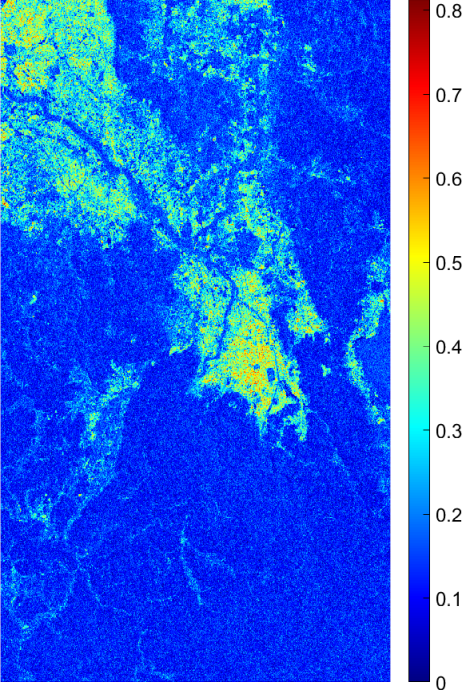}} \\
	\caption{(a), (b) and (c) represent the Pauli RGB image, $\delta_{13}$ and $\delta_{23}$ colormaps for the AIRSAR L-band data over San Francisco, USA respectively. (d), (e) and (f) represent the Pauli RGB image, $\delta_{13}$ and $\delta_{23}$ colormaps for the ALOS-2 L-band over Kyoto, Japan respectively.}
	\label{fig:SF_KY_deltas}
\end{figure}

The $\delta_{13}$ and the $\delta_{23}$ parameters for the San Francisco (AIRSAR L-Band), and the Kyoto (ALOS-2 L-Band) images are shown in Figure~\ref{fig:SF_KY_deltas}(a)-(c) and in Figure~\ref{fig:SF_KY_deltas}(d)-(f) respectively. It can be clearly seen that the areas with buildings and vegetation are well discriminated in both these images. The values over areas of vegetation are very low for both the $\delta_{ij}$ values ranging from 0 to 0.2. \textcolor{red}{This is desirable as it prevents the under-estimation of the volume power from forested areas. The low $\eta_{i}$ values further reduce this compensation. For a pure volume scatterer theoretically no compensation is performed.} On the other hand, in the urban areas, the $\delta_{23}$ is dominant over $\delta_{13}$ values, the former ranging between 0.5 to 0.7 while the later varies from 0 to 0.4. This indicates a strong coupling between the $(S_{\mathrm{HH}} - S_{\mathrm{VV}})$ and the $(S_{\mathrm{HV}})$ components in the urban areas. 

The parameters $\delta_{13}$ and the $\delta_{23}$ are quantitatively analyzed over the two rectangular regions: A and B as shown in Figure~\ref{fig:SF_KY_deltas}(a) and  Figure~\ref{fig:SF_KY_deltas}(d). It can be seen that the values of the $\delta_{13}$ parameter $(0.16\pm0.20)$ and $(0.19\pm0.22)$ is lower than the $\delta_{23}$ parameter $(0.37\pm0.31)$ and $(0.41\pm0.30)$ for the rectangles A and B (size 100$\times$100 pixels) respectively for the San Francisco data. Similar trend is also observed for the Kyoto data with $\delta_{13}$ parameter $(0.14\pm0.15)$ and $(0.12\pm0.13)$, which is lower than the $\delta_{23}$ parameter $(0.44\pm0.21)$ and $(0.41\pm0.21)$ for the rectangles A and B respectively. The high standard deviation in both the parameters indicates that the difference of the co- and cross-polarization correlation is spread over wide ranges. In urban environment, this attribute is quite likely due to complex scattering at longer wavelengths.  

\begin{figure}[!hbt]
	\centering
	\subfigure[]{\adjincludegraphics[trim={0 0 {0.13\width} 0},clip, width=0.32\textwidth]{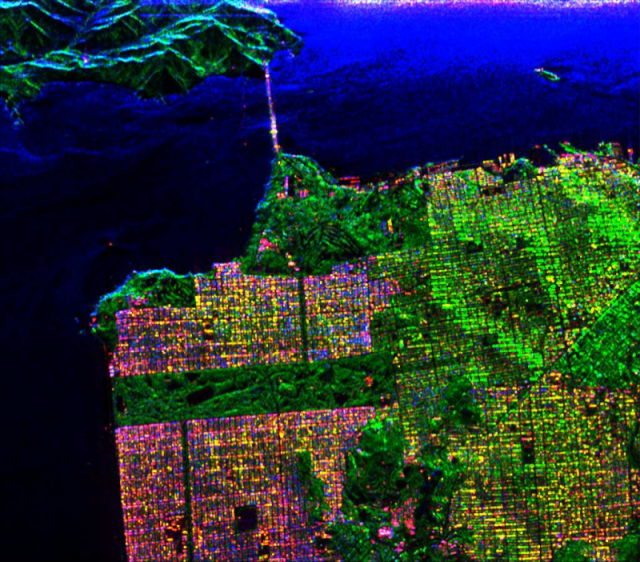}} 
	\subfigure[]{\adjincludegraphics[trim={0 0 {0.13\width} 0},clip, width=0.32\textwidth]{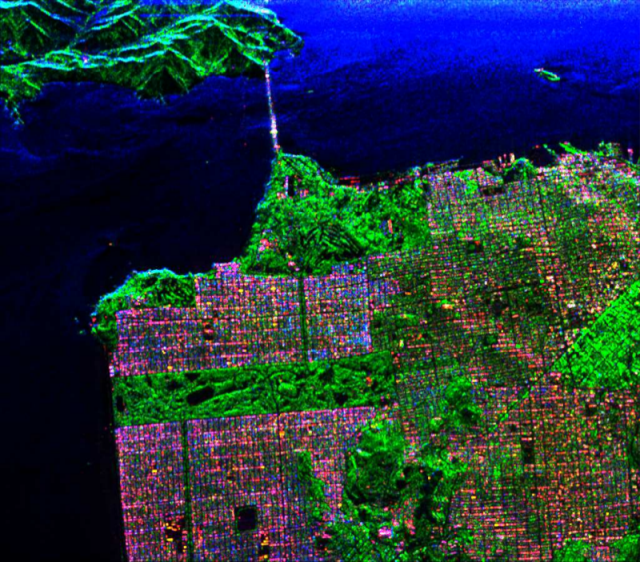}}  
	\subfigure[]{\adjincludegraphics[trim={0 0 {0.13\width} 0},clip, width=0.32\textwidth]{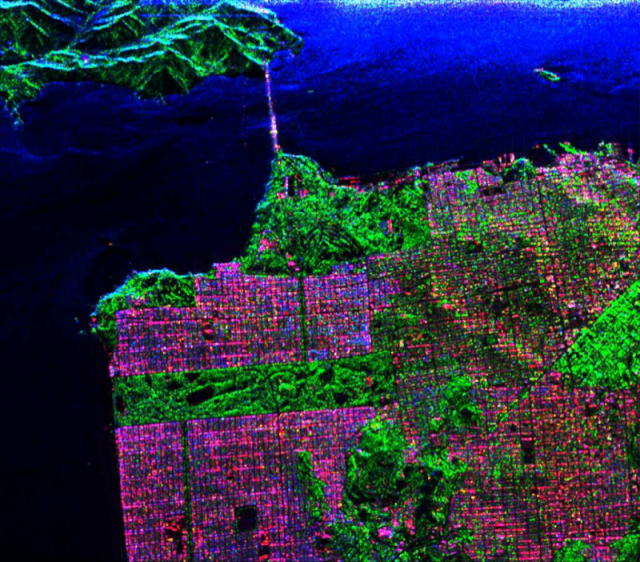}} \\
	\subfigure[]{\includegraphics[width=0.25\columnwidth]{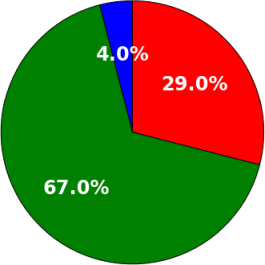}} \hspace{10mm}
	\subfigure[]{\includegraphics[width=0.25\columnwidth]{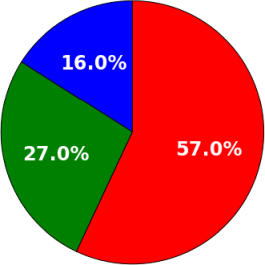}} \hspace{10mm}
	\subfigure[]{\includegraphics[width=0.25\columnwidth]{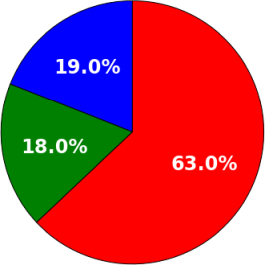}}\\
	\subfigure[]{\includegraphics[width=0.25\columnwidth]{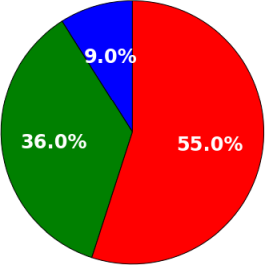}} \hspace{10mm}
	\subfigure[]{\includegraphics[width=0.25\columnwidth]{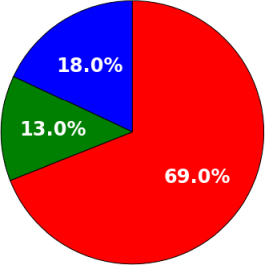}} \hspace{10mm}
	\subfigure[]{\includegraphics[width=0.25\columnwidth]{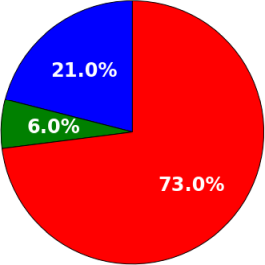}}\\	
	\caption{AIRSAR L-band San Francisco image: (a) Y4O (b) Y4R (c) RD--Y4O; (d), (e) and (f) are corresponding scattering powers for rectangle A; (g), (h) and (i) are scattering powers for rectangle B.}
	\label{fig:SF_airsar}
\end{figure}

The RD measures, $\zeta_{1}$ and $\zeta_{2}$ which are used to modify the powers in the proposed decomposition are analyzed for the two rectangular areas A and B respectively for the San Francisco and the Kyoto image. The scattering powers obtained with the proposed decomposition (RD--Y4O) are compared with the Y4O and the Y4R. It can be observed in Figure~\ref{fig:SF_airsar}(c) that the double-bounce scattering obtained from RD--Y4O in urban areas are enhanced compared to the Y4O and Y4R shown in Figure~\ref{fig:SF_airsar}(a) and Figure~\ref{fig:SF_airsar}(b) respectively. The vegetation areas and the water surface seem more or less same in all the three decompositions. 

The scattering power pie chart for the San Francisco image is shown for the rectangular region A in \ref{fig:SF_airsar}(d)--(e)--(f). The double-bounce scattering power $(P_{\mathrm{d}} \;\mathrm{or}\; \overset{\sim}{{P}}_{\mathrm{d}})$ is shown in red, surface scattering power $(P_{\mathrm{s}} \;\mathrm{or}\; \overset{\sim}{{P}}_{\mathrm{s}})$ in blue, and volume scattering power $(P_{\mathrm{v}} \;\mathrm{or}\; \overset{\sim}{{P}}_{\mathrm{v}})$ in green respectively. It can be seen that for urban area A, the double bounce power $\overset{\sim}{{P}}_{\mathrm{d}}$ using RD--Y4O has increased by more than double than the Y4O $P_{\mathrm{d}}$ (\emph{i.e.,} from 29\% to 63\%), exceeding the corresponding $P_{\mathrm{d}}$ power in Y4R by 6\% (\emph{i.e.,} from 57\% to 63\%). The volume power $P_{\mathrm{v}}$ has decreased from 67\% for Y4O to 27\% for Y4R to 18\% of $\overset{\sim}{{P}}_{\mathrm{v}}$ for RD-Y4O. The surface power $\overset{\sim}{{P}}_{\mathrm{s}}$ has also increased by 3\% from the Y4R model. \textcolor{red}{It can be noted that the AGU--DOP method corrects the POA. However, some buildings show volume scattering power which may be due to critical azimuthal slopes indigenous to the region A.} For the region B which also lies within an urban settlement, the scattering power pie chart is shown in Figure~\ref{fig:SF_airsar}(g)--(h)--(i). The double-bounce power $\overset{\sim}{{P}}_{\mathrm{d}}$ has increased to 60\% from 42\% for Y4O and 54\% for Y4R. The surface power $\overset{\sim}{{P}}_{\mathrm{s}}$ has also increased by 9\% from Y4O. 

\begin{figure*}[!hbt]
		\centering
		\subfigure[]{\adjincludegraphics[trim={0 {0.6\width} 0 0},clip, width=0.32\textwidth]{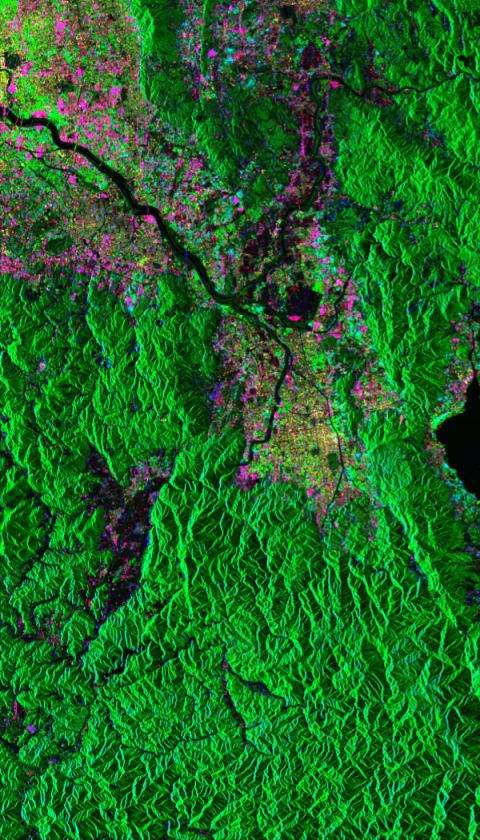}} 
		\subfigure[]{\adjincludegraphics[trim={0 {0.6\width} 0 0},clip, width=0.32\textwidth]{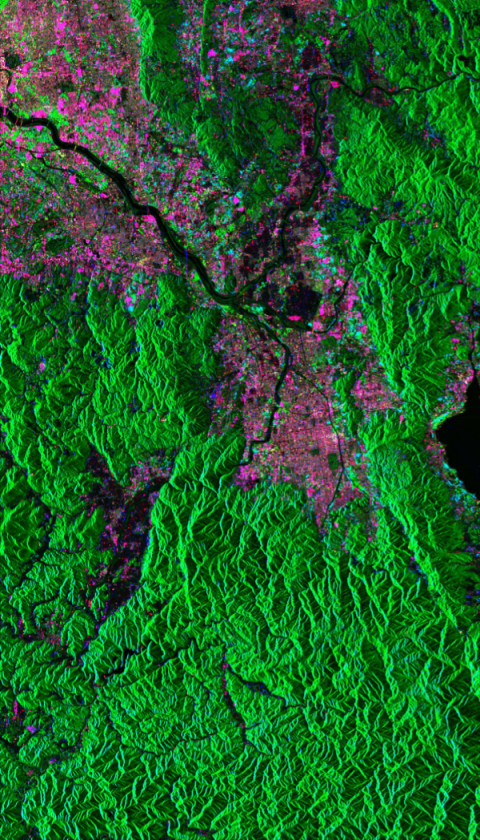}}  
		\subfigure[]{\adjincludegraphics[trim={0 {0.6\width} 0 0},clip, width=0.32\textwidth]{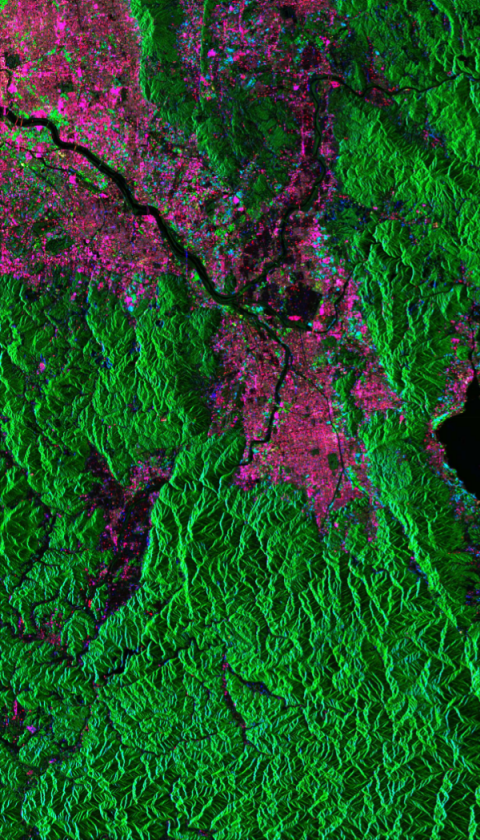}} \\
		\subfigure[]{\includegraphics[width=0.25\columnwidth]{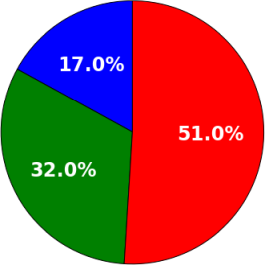}} \hspace{10mm}
		\subfigure[]{\includegraphics[width=0.25\columnwidth]{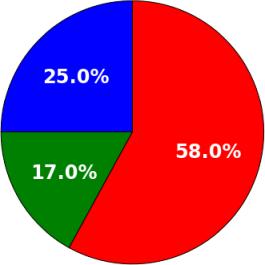}} \hspace{10mm}
		\subfigure[]{\includegraphics[width=0.25\columnwidth]{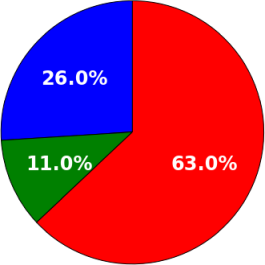}}\\
		\subfigure[]{\includegraphics[width=0.25\columnwidth]{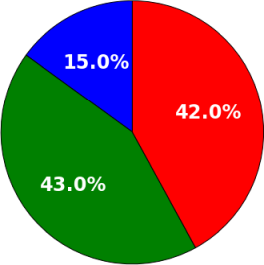}} \hspace{10mm}
		\subfigure[]{\includegraphics[width=0.25\columnwidth]{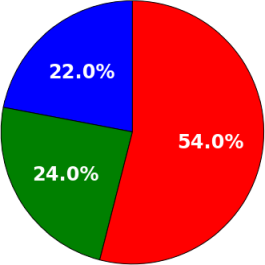}} \hspace{10mm}
		\subfigure[]{\includegraphics[width=0.25\columnwidth]{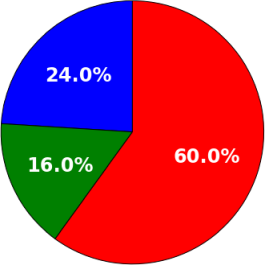}}\\	
		\caption{ALOS-2 L-band Kyoto image: (a) Y4O (b) Y4R (c) RD--Y4O; (d), (e) and (f) are corresponding scattering powers for rectangle A; (g), (h) and (i) are scattering powers for rectangle B.}
		\label{fig:Kyoto_ALOS2}
\end{figure*}

The scattering power pie chart for the Kyoto image is shown for the rectangular region A in \ref{fig:Kyoto_ALOS2}(d)--(e)--(f) and for the rectangle B in \ref{fig:Kyoto_ALOS2}(g)--(h)--(i). The double bounce power $\overset{\sim}{{P}}_{\mathrm{d}}$ using RD--Y4O has increased to 63\% from 58\% in the Y4R and 51\% in the Y4O for the region A. Similarly, the $\overset{\sim}{{P}}_{\mathrm{d}}$ power has increased to 60\% from 54\% in the Y4R and 42\% in the Y4O for the region B. Even though the surface powers $\overset{\sim}{{P}}_{\mathrm{s}}$ has increased by 9\% for both the areas A and B from the Y4O, there is negligible increase in this power from the Y4R. In both cases, \emph{i.e.,} for San Francisco and Kyoto datasets, the enhancement of powers for coherent targets indicate that the information content in the off-diagonal elements $(T_{13})$ and $(T_{23})$ of the coherency matrix are vital. In fact, larger enhancements can be associated with larger orientation angles (\emph{i.e.,} azimuth slopes).  
Table~\ref{table:negative_powers} shows the proportion of pixels with negative powers. It can be observed that by the RD--Y4O, the proportion of pixels with negative powers show a decline which is one of the criteria to judge a decomposition model. For the AIRSAR San Francisco image, the value is reduced from 13.8\% in Y4O to 5.3\% in RD--Y4O. Similarly, in ALOS-2 Kyoto image also the minimum \% among the three decompositions i.e. 8.5\% is achieved with the proposed decomposition.
\begin{table}[!hbt]
		\caption{\textcolor{red}{Proportion of pixels with negative powers (\%)}}
		\centering
		\begin{tabular}{|c || c| c| c|}
		\hline
		\textcolor{red}{Data Set} & Y4O & Y4R & RD-Y4O \\ \hline
		AIRSAR & 13.8 & 7.6 & 5.3 \\ \hline
		ALOS-2 & 17.8 & 15.0 & 8.5\\ \hline
		\end{tabular}
		\label{table:negative_powers}
\end{table}

\section{Conclusion}
In this letter, a methodology is proposed for utilizing the information contained in the off-diagonal elements of the coherency matrix for improving of PolSAR decomposition scattering powers. The enhancement in scattering powers is observed only in desired regions e.g. where azimuth slope effects are present. Negative power pixels have been reduced. The integration with the established DOP optimization criterion $(\mathrm{AGU-DOP})$, is a good amalgam of physics of scattering with a statistical quantity as correlation in our proposed method. In the future, the concept of relative decorrelation may be extended to applications as enhanced classification, change detection and segmentation. 
\textcolor{red}{\section*{Acknowledgments}
	The authors would like to thank Mr. Abhishek Maity and Mr. Shaunak De for assisting with the technical editing.}

\section*{Disclosure statement}
No potential conflict of interest was reported by the authors.
\bibliographystyle{tRSL}
\bibliography{mybibfile}

\end{document}